\documentclass[pra,twocolumn,10pt,showpacs,preprintnumbers,amsmath,amssymb,aps]{revtex4-1}

\usepackage{graphicx}

\begin{document}

\title{Efficient Collection of Single Photons Emitted from a Trapped Ion into a Single Mode Fiber for Scalable 
			Quantum Information Processing}

\author{Taehyun Kim, Peter Maunz, and Jungsang Kim}
\affiliation{Fitzpatrick Institute for Photonics, Electrical and Computer Engineering Department, Duke University,
Durham, North Carolina 27708, USA}

\begin{abstract}
Interference and coincidence detection of two photons emitted by two remote ions
can lead to an entangled state which is a critical resource for scalable quantum information processing.
Currently, the success probabilities of experimental realizations of this protocol are mainly limited by 
low coupling efficiency of a photon emitted by an ion into a single mode fiber.
Here, we consider two strategies to enhance the collection probability
of a photon emitted from
a trapped Yb$^+$ ion, using analytic methods that can be easily applied to other types of ion or neutral atoms.
Our analysis shows that we can achieve fiber coupling efficiency of over 30~\%
with an optical cavity made of a flat fiber tip and a spherical mirror.
We also investigate ways to increase the fiber coupling efficiency using high numerical aperture optics,
and show that collection probability of over 15~\%  is possible with proper control of aberration. 
\end{abstract}


\maketitle
\section{INTRODUCTION}
\label{Section:Introduction}
Entanglement is a unique feature of quantum physics and
it has found various applications ranging from the experimental verification of the foundation of quantum physics \cite{EPRParadox}
to many quantum information protocols such as quantum teleportation \cite{QTeleportation},
quantum cryptography \cite{QCryptography}, quantum computation \cite{QTeleportationBasedComputing,ClusterStateComputing},
and long distance quantum communication \cite{QRepeater}.
Most of these applications require a large number of entangled states distributed over long distances,
which can be generated using one of two approaches.
In the first approach, one can locally generate a pair of entangled photons \cite{PhotonEntanglementSource},
distribute them to remote sites, and map these flying qubits into stationary (memory) qubits \cite{QubitTransfer}.
To efficiently transfer a flying qubit into a stationary qubit, 
the spectral and spatial modes of the flying qubit have to be matched
to the time-inverted emission properties of the memory qubit, which is the most significant challenge in this approach.
In the second approach, 
one first generates flying qubits that are entangled with a local stationary qubit at each site,
sends these photons to a common location, and then interferes these two photons using a 50:50 beam splitter \cite{DLCZ, RemoteEntanglementTheory}.
With 50 \% chance, these photon pairs are projected into one of two distinguishable Bell states,
and once one of these two Bell states is detected,
the two stationary qubits at the remote sites are also projected to a maximally entangled Bell state \cite{EntanglementSwapping}.
The advantage of this approach is that the spectral properties of the two photons 
are naturally matched if one uses the same type of memory qubits.
In 
order to achieve interference of the two photons with high visibility,
the photons have to be indistinguishable.
Therefore, we need to collect emitted photons into a single spatial mode, such as a single mode fiber.
Several experiments demonstrated establishment of entanglement between remote trapped ions
using this approach \cite{RemoteEntanglement,RemoteEntanglementPolarization,QuantumTeleportationExperiment}.
While these experiments realized 
high fidelity 
entangled states, the success probability of the probabilistic entanglement process is very small.
The success probability is mainly limited by the small photon collection probability of about 0.4~\%
given by
the small numerical aperture (NA) of the collection optics and 
an insufficient overlap between the spatial mode of the single mode fiber and the emission pattern of an ion.
Because this protocol requires one to collect photons from both sites and observe a coincidence,
the overall success probability scales quadratically with the single fiber coupling efficiency,
and was limited to only about $2\cdot10^{-8}$ (see Table \ref{Tab:GenerationRate} in Sec. \ref{Discussion}) 
\cite{QuantumTeleportationExperiment}.
To apply this system for large scale quantum information processing or long distance quantum communication, 
it is critical to dramatically enhance the collection probability of photons into a single mode fiber.

In this paper, we show systematic approaches to increase the fiber coupling efficiency of a single photon 
while preserving ion-photon entanglement.
We will analyze the design of an optimized optical cavity
that can be integrated with a surface ion trap,
and evaluate various ways to achieve high NA collection optics compatible with surface traps 
to enhance light collection into a single mode fiber.

\begin{figure}
	\begin{center}
		\includegraphics[width=8cm]{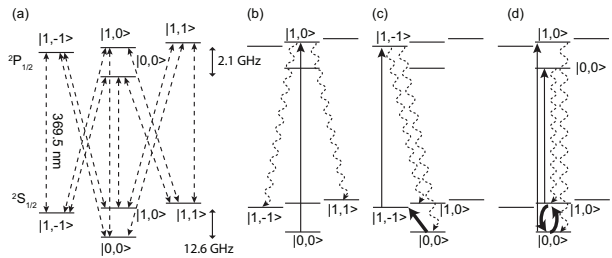} 
	\end{center}
	\caption{Ion-photon entanglement schemes in $^{171}{\rm Yb}^+$.
	(a) Simplified level scheme of $^{171}{\rm Yb}^+$ with dipole-allowed transitions.
	Zeeman levels are denoted by $|F,m_F\rangle$, where $F$ is the total angular momentum including nuclear spin
	and $m_F$ is its projection onto the quantization axis.
	$^2S_{1/2}|1,0\rangle$ and $^2S_{1/2}|0,0\rangle$ are 
	first-order magnetic field-insensitive hyperfine levels (clock states).
	(b) When the ion decays from $^2P_{1/2}|1,0\rangle$ 
	to one of the Zeeman levels in $^2S_{1/2}|1,\pm 1\rangle$,
	the polarization of the emitted photon gets entangled with two Zeeman levels \cite{RemoteEntanglementPolarization}.
	(c) The frequency of the emitted photon and the atomic clock states are entangled after a decay from $^2P_{1/2}|1,-1\rangle$
	\cite{RemoteEntanglement}.
	(d) Due to the selection rules, excitation of $^{171}{\rm Yb}^+$ prepared 
	in a superposition of the clock states in $^2S_{1/2}$ generates
	a coherent superposition of the excited states (two $^2P_{1/2}$ states with $m_F = 0$).
	Once excited state decays back to the original state,
	the frequency of the emitted photon is entangled with	clock states \cite{QuantumTeleportationExperiment}.
	In $^{171}{\rm Yb}^+$ system, $^2S_{1/2}|0,0\rangle$ is generally used as an initial state 
	due to its high preparation fidelity. Therefore (c) and (d) scheme require additional microwave or optical Raman transitions
	denoted by thick arrows.
	}
	\label{Fig:EntanglementScheme}
\end{figure}

Our analysis mainly focuses on $^{171}{\rm Yb}^+$ whose simplified level scheme is shown in Figure \ref{Fig:EntanglementScheme} (a) \cite{YbSystem}, 
but the conclusions can be applied to other atomic systems.
There are several ways to entangle a photon's degree of freedom with internal electronic states of a trapped $^{171}{\rm Yb}^+$ ion. 
If Zeeman levels are denoted by $|F,m_F\rangle$, where $F$ is the total angular momentum
	and $m_F$ is its projection onto the quantization axis,
one can excite the Yb ion to $^2P_{1/2}|1,0\rangle$ and collect the photon from the $\sigma^\pm$ transition 
as shown in Figure \ref{Fig:EntanglementScheme} (b).
The polarization of the photon will then be entangled 
with the two ground state Zeeman levels $^2S_{1/2}|1,\pm1\rangle$ \cite{RemoteEntanglementPolarization}.
Alternatively, one can collect a photon generated by $\sigma$-transitions from $^2P_{1/2}|1,-1\rangle$ as shown in Figure \ref{Fig:EntanglementScheme} (c) 
or by $\pi$-transitions from a superposition of $^2P_{1/2}|1,0\rangle$ and $^2P_{1/2}|0,0\rangle$
to the atomic clock states composed of $^2S_{1/2}|1,0\rangle$ and $^2S_{1/2}|0,0\rangle$
as shown in Figure \ref{Fig:EntanglementScheme} (d).
The frequency of the photon is then entangled with the clock state qubits.
The spatial emission pattern of photons from $\sigma^\pm$ and $\pi$-transitions 
and their spectral properties together with
the ability to filter out photons from unwanted transitions
will dictate strategies that can be adopted to enhance the photon collection efficiency.
For remote entanglement generation, we will analyze the approach using an optical cavity in Section \ref{CavitySection},
and an approach using high NA optics (either a reflector or a lens) in Section \ref{HighNASection}. Section \ref{Discussion} estimates the entanglement generation rate we can achieve based on these approaches, and conclusions are presented in Section \ref{Conclusions}.

\section{Cavity enhanced photon collection}
\label{CavitySection}
A small high finesse cavity can be used to collect light from a single atom
by induced emission into the cavity mode.
Although efficient light collection using an optical cavity has been demonstrated
for neutral atoms \cite{SinglePhotonSourceWithCs, SinglePhotonSourceWithRb},
additional challenges have to be met for trapped ions:
readout transitions of trapped ions are typically in the ultraviolet (UV)
and the ion trap has to fit in a small cavity with UV-dielectric coatings.
In addition, trapped ions have to be shielded from the charges which can accumulate on dielectric surfaces \cite{DielectricCharging}.

\begin{figure}
	\begin{center}
		\includegraphics[width=4cm]{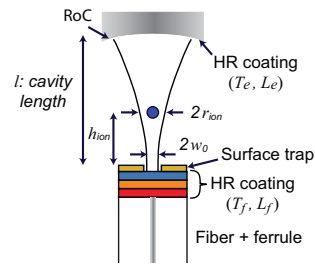} 
	\end{center}
	\caption{Geometry of an optical cavity integrated with surface trap. 
	Optical cavity is composed of a small spherical mirror
	and a fiber in a ferrule whose tip is coated with high reflecting dielectric layers.
	$l$ is cavity length, $w_0$ is beam waist at the fiber tip,
	and $r_{ion}$ is beam radius at the trapping position of the ion.
	Surface trap can be patterned with Au on the fiber/ferrule surface whose diameter is $>$2~mm.}
	\label{Fig:FiberCavity}
\end{figure}

The collection probability of a single photon from an excited atom into a single mode fiber
is dictated by several factors.
First, the excitation has to leave the coupled atom-cavity system
via the cavity decay channel rather than spontaneous emission, characterized by the photon extraction probability $P_{cavity}$.
While some photons leave the cavity through the desired output channel,
other photons are either lost due to scattering and absorption in the mirrors
or transmitted through the undesired mirror(s).
We define $r_t$ as the fraction of the desired transmission over the total cavity decay.
Finally, we consider overlap between the mode of the cavity and that of the single mode fiber,
which will be called mode matching factor $\epsilon$.
The overall probability $P_{fiber}$ of extracting the photon through the single mode fiber is then given by
\begin{equation}
P_{fiber} = \epsilon \cdot r_t \cdot P_{cavity}. \label{Eq:P_fiber}
\end{equation}
In this section, we will investigate how to maximize each factor within the physical constraints.
We will mainly consider a cavity geometry shown in Figure \ref{Fig:FiberCavity} \cite{IntegratedOptics}
and optimize system parameters to maximize the collection efficiency
under  the given constraints.
This discussion will also identify components that deserve further investigation 
to improve the performance of the system in the future, and provide a metric to compare different cavity systems.

When an atomic system is coupled to a cavity mode,
a single photon can be generated  
either by fast excitation with resonant $\pi$-pulse followed by spontaneous emission \cite{RemoteEntanglement,RemoteEntanglementPolarization,QuantumTeleportationExperiment,FastExcitation}, or by stimulated Raman adiabatic passage (STIRAP) process \cite{SinglePhotonSourceWithCs, SinglePhotonSourceWithRb}.
In this section, we will also compare these two methods in terms of collection probability $P_{fiber}$
and repetition rate.

\subsection{Fast excitation with resonant $\pi$-pulse: Design principles}
\label{FastExcitation}
In this subsection, we consider the case where the ion is excited by an ultrafast pulse 
whose duration is short compared to any time constants of the system, such that the initial state is an excited state of the ion.
When an excited ion is placed inside an optical cavity, 
the spontaneous emission rate into the cavity mode
can be enhanced by the Purcell factor, $2C+1$ \cite{PurcellEnhancement},
where $C=g^2/2\kappa\gamma$ is the cooperativity, $g$ is the coherent ion-field coupling constant,
$\kappa$ is the cavity-field decay rate, and $2\gamma$ is the spontaneous emission rate.

The parameter $g^2$ is inversely proportional to the cavity mode volume 
and $\kappa = \pi \cdot FSR/\mathcal{F}$, where $FSR$ and  $\mathcal{F}$ are the free spectral range and the finesse of the cavity, respectively.
Therefore, we need a high finesse cavity with small mode volume to achieve large cooperativity.
Small mode volume cavities can be realized by many different approaches 
such as microdisk \cite{MicroDisk}, microsphere  \cite{MicroSphere}, microtoroid  \cite{MicroToroid}, and photonic crystals  \cite{PhotonicCrystal}.
Integration with ion traps requires a cavity with empty mode volume, 
so we will consider only traditional open cavities formed by two (curved) mirrors. 
Cavity mode shape between two spherical mirrors is Gaussian and the 
mode volume is $\pi w_0^2 l/4$,
where $w_0$ is the beam waist of the Gaussian mode function 
and $l$ is length of the cavity.
The cooperativity is reduced to $C = 3 \mathcal{F}\cdot \text{Br}\cdot \lambda^2/(\pi^3 r_{ion}^2)$, where Br is a branching ratio,
$\lambda$ is wavelength of transition, and $r_{ion}$ is the radius of the mode function at ion location.
In this system, we need high $\mathcal{F}$ and small $r_{ion}$ to maximize $C$.

Using these cavity parameters, one can express the standard formula for $P_{cavity}$ \cite{CavityProbability} as a function of the cavity length ($l$)
and the beam radius at ion location ($r_{ion}$):
\begin{eqnarray}
P_{cavity} & = & \left(\frac{\kappa}{\kappa+\gamma}\right)\left(\frac{2C}{2C+1}\right) \nonumber \\
& = & \left(\frac{1}{1+l/l_c}\right)\left(\frac{1}{1+r_{ion}^2/r_c^2}\right),
\label{Eq:Pcavity}
\end{eqnarray}
where $l_c$ is a characteristic cavity length defined as $l_c = \pi c/(2 \gamma \mathcal{F})$
and $r_c$ is a characteristic beam radius defined as $r_c = \sqrt{6\lambda^2\mathcal{F}\cdot \text{Br}/\pi^3}$.
Eq. (\ref{Eq:Pcavity}) implies that if either $l$ or $r_{ion}$
is much larger than the corresponding characteristic lengths,
the overall $P_{cavity}$ will decrease rapidly. The characteristic cavity length (beam radius) decreases (increases) as
the finesse of the cavity increases. This indicates that for high finesse cavities 
one should use a short cavity but the beam radius can be larger,
and the design shifts to longer cavities with smaller beam radius as the finesse of the cavity decreases.

Cavity finesse $\mathcal{F}$ is defined by $2\pi\sqrt{1-\mathcal{L}}/\mathcal{L}$,
where $\mathcal{L}=T_f + T_e + L_f + L_e$ is the total loss of the cavity,
$T_f$ and $L_f$ ($T_e$ and $L_e$) are transmission and passive loss through dielectric coating on the fiber tip (the other end mirror) in Figure \ref{Fig:FiberCavity}, respectively.
Passive loss is due to scattering loss mainly determined by surface roughness \cite{FiberCavity}
and absorption in the dielectric coating.
We can define $\mathcal{F}_0 \equiv 2\pi\sqrt{1-(L_f+L_e)}/(L_f+L_e)$,
to denote the maximum achievable finesse limited by the uncontrollable passive loss.
Then the actual finesse can be approximated to $\mathcal{F} \simeq \mathcal{F}_0(1-r_t)$ where $r_t = T_f/\mathcal{L}$.
Assuming the mode matching factor $\epsilon$ is constant,
one can see that $P_{fiber}$ defined in Eq. (\ref{Eq:P_fiber}) is maximized when
\begin{equation}
r_t = r_{t0} \equiv  \frac{1}{1+1/\sqrt{(1+2C_0)(1+l/l_{c0})}}
\label{Eq:r_t}
\end{equation}
to a value
\begin{equation}
P_{fiber,max}=\frac{C_0}{1+C_0+l/2l_{c0}+\sqrt{(1+2C_0)(1+l/l_{c0})}}, 
\label{Eq:MaximizedPfiber}
\end{equation}
where $C_0 \equiv 3\mathcal{F}_0\cdot \text{Br}\cdot \lambda^2/(\pi^3 r_{ion}^2)$ and $l_{c0} \equiv \pi c/(2 \gamma \mathcal{F}_0)$ are the cooperativity and characteristic length with maximum possible finesse $\mathcal{F}_0$, respectively. From Eq. (\ref{Eq:MaximizedPfiber}), one can see that $C_0$ should be maximized while $l$ should be minimized in order to maximize the fiber coupling efficiency.

Finally, fiber collection efficiency is further reduced 
by  mismatch between cavity mode ($E$) and fiber mode ($G_{fiber}$) at the fiber tip characterized by the mode matching factor
\begin{equation}
\epsilon=
\frac{|\int_A da \vec{E}^*\cdot \vec{G}_{fiber}|^2}
	{\int_A da |\vec{E}|^2 \int_A da |\vec{G}_{fiber}|^2},
\label{Eq:ModeMatching}
\end{equation}
where $G_{fiber}$ and $E$ are the transverse Gaussian mode shapes of fiber and cavity, respectively.
The optimal $w_0$ of the fiber cavity should be determined to maximize the coupling $P_{fiber}$ to an ion at height $h_{ion}$, which may not necessarily match with the fiber mode. If one uses additional optics to couple the cavity mode into a fiber mode rather than integrate the fiber mode directly underneath the flat cavity mirror, the mode matching factor given by Eq. (\ref{Eq:ModeMatching}) can be improved substantially, approaching a value of 1.

\subsection{Cavity design example}
\label{NumericalExamples}
The optimum design of the cavity-ion trap system is often constrained by the ability to achieve high performance components. 
In our system, the dominating constraint is the finesse $\mathcal{F}_0$  and the radius of curvature (RoC) for the curved cavity mirror one can achieve, limited by the coating quality and polishing capability.
Compared to near-infrared wavelengths used in neutral atom experiments \cite{SinglePhotonSourceWithCs, SinglePhotonSourceWithRb},
the passive loss of cavity mirror coating is significantly higher in the ultraviolet range relevant for most trapped ions.
Scattering loss from a rough surface is given by $1-\exp[-(4 \pi \sigma/\lambda)^2]$ \cite{FiberCavity} where $\sigma$ is RMS surface roughness.
For example, we observed $\approx$ 150 parts per million (ppm) loss at 369.5~nm on a super-polished substrate
whose RMS surface roughness is measured to be $<$0.1~nm, corresponding to estimated scattering loss of $<$12 ppm. We can attribute most of
the loss in this case to absorption loss in the coating material in the UV wavelength range.
Since the RMS surface roughness of the fiber tip and the mirror substrate with small RoC
is generally larger than that of a super-polished mirror substrate with large RoC,
it is reasonable to assume that the uncontrolled loss in our mirrors is  $L_f + L_e \approx$1500~ppm, corresponding to $\mathcal{F}_0 \approx 4200$.

Next, we have to pick the radius of the cavity mode $r_{ion}$ at the ion location and the cavity length $l$. In order to maximize $C_0$, one should find a minimum value of $r_{ion}$.
Since the ion cannot be trapped at the beam waist  in a fiber tip cavity shown in Fig. (\ref{Fig:FiberCavity}),
the optimal waist $w_0$ is determined by the distance between the ion and the waist which is located on the flat fiber tip.
When this distance is given by $h_{ion}$, 
$r_{ion}$ can be minimized if $w_0 = \sqrt{h_{ion}\cdot\lambda/\pi}$.
In a surface trap, the minimum ion height from the surface is constrained by considerations such as
the exposed dielectric material \cite{DielectricCharging}, anomalous heating \cite{AnomalousHeating}
and the scattering of laser light from the surface, and $h_{ion}$ = 50~$\mu$m is a good compromise.
In this case, the optimal waist is $w_0$ = 2.4~$\mu$m, corresponding to $r_{ion}$ = 3.4~$\mu$m.

For a cavity created between a flat mirror and a concave mirror, the beam waist of the Gaussian mode is located at the flat mirror surface. The length of the cavity $l$ is related to the Rayleigh range $z_R=\pi w_0^2/ \lambda$ of the Gaussian mode as $z_R^2=({\rm RoC}-l)l$. 
To trap the ion inside the cavity volume, we require $l>h_{ion} = z_R$ and therefore $l$ should be chosen to be very close to the RoC. 
A practical mirror one can polish using conventional super-polishing technique has a RoC~$\geq$~5~mm, 
so we choose $l\approx {\rm RoC}-z_R^2/{\rm RoC} \approx$ 5~mm. The cavity length must be aligned to be about 0.5~$\mu$m shorter than the RoC of the mirror. Laser ablation can create a smooth curved surface with RoC between 40~$\mu$m and 2~mm 
with 0.2~nm RMS roughness \cite{LaserAblation}, but the aperture diameter of a mirror fabricated in this way is limited to $< 100~\mu$m 
which is too small for near-concentric cavity geometry.

\begin{figure}
	\begin{center}
		\includegraphics[width=8cm]{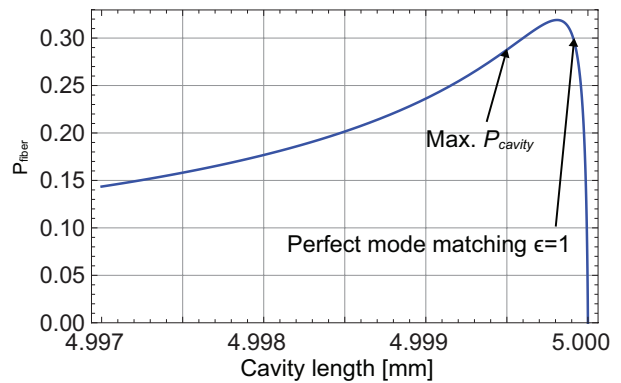} 
	\end{center}
	\caption {Probability of collecting photons with a fiber cavity into the mode of a single mode fiber
	as a function of cavity length ($l$).
	As the cavity length approaches the RoC of the spherical mirror, 
	the beam waist at the fiber tip shrinks.
	$T_f$ = 9,500~ppm and $T_e+L_f+L_e$= 1500~ppm are assumed.
	The radius $r_{ion}$ is smallest for a cavity length of 4.9995~mm 
	while best mode overlap between cavity and fiber is achieved for a length of 4.9999 mm.}
	\label{Fig:FiberOverlap}
\end{figure}

Using these parameters, we obtain $C_0 \approx 4.8$ and $l_{c0} \approx$ 1.8~mm. 
The transmission of the out coupling mirror is determined by setting $r_t \approx 86 \%$ according to Eq. (\ref{Eq:r_t}), corresponding to $T_f \approx$ 9,500 ppm. 
This choice determines critical cavity parameters as $\mathcal{F} =  570$, $C = $ 0.65, $l_c$=13~mm and $r_c = $ 3.9~$\mu$m. 
Measured beam waist of a typical UV fiber at 369.5~nm is about 1.5~$\mu$m and
in Figure \ref{Fig:FiberOverlap} the change of the $P_{fiber}$ as a function of the cavity length $l$
for an ion height $h_{ion}$ = 50~$\mu$m is shown.
The mode matching efficiency $\epsilon$ is about 82~\% if $w_0$ is optimized for maximum $P_{cavity}$
($w_0$ = 2.4~$\mu$m), and $P_{cavity}$ is reduced when $\epsilon$ is 100 \%.
Therefore, there is an optimal length that maximizes $P_{fiber}$, near which
one can achieve more than 30~\% of coupling efficiency compared to 0.4~\% \cite{QuantumTeleportationExperiment}.

If additional optics is used to couple the cavity mode into a single mode fiber (instead of using fiber tip as the flat mirror of the cavity), we can achieve $\epsilon \approx$1 and a maximum fiber coupling efficiency of over 35 \%. A comparison with a cavity constructed using two concave mirrors is presented in Appendix \ref{Appendix:A}.

In summary, we expect to reach $(g, \kappa, \gamma)/2\pi \approx (18, 26, 10)$ MHz using a fiber tip cavity.
With pulsed excitation, one can achieve more than 30 \% coupling efficiency into the single mode fiber
if the minimum available RoC is about 5~mm,
and the average rate of  single photon emission process is on the order of $\text{max}(1/g,1/\kappa)$.

\subsection{Excitation by stimulated Raman adiabatic passage}
\label{STIRAP}
In an atomic system where the initial state $|i\rangle$ and 
the final states $|f_1\rangle$, $|f_2\rangle$ are different
as shown in Fig. \ref{Fig:EntanglementScheme} (b)
and can be connected by a resonantly enhanced Raman transition,
the STIRAP process can be used to generate a single photon 
in the cavity mode \cite{SinglePhotonSourceWithCs, SinglePhotonSourceWithRb}.

For this process, a laser beam is resonant with one branch of the Raman transition ($|i\rangle \Leftrightarrow |e\rangle$) with Rabi frequency $\Omega_P$,
and the cavity vacuum stimulates the emission of the photon on the other branch ($|f\rangle \Leftrightarrow |e\rangle$)
with vacuum Rabi frequency $2g$.
As $\Omega_P$ is slowly increased from zero, the ion adiabatically follows the dark state of the Hamiltonian starting 
from $|i\rangle$ to $|f\rangle$.
Because the excited state $|e\rangle$ is not populated, spontaneous emission does not occur.
The photon collection probability $P_{cavity}$ is $2C/(2C+1)$.
If used together with the fiber cavity geometry discussed above, $P_{cavity} \approx 0.56$ and $P_{fiber} \approx 0.40$ can be reached.

If one is interested in an ideal single photon source, STIRAP is a better choice because of its larger photon collection efficiency per trial.
Furthermore, in contrast to the fast excitation, the pulse shape of the emitted photon can be controlled by $\Omega_P(t)$.
For the protocol to generate entangled ion pair based on entanglement swapping between a photon pair,
successful generation rate of the photon pair is the relevant figure of merit,
and is proportional to $P_{fiber}^2 \cdot R_{rep}$, where $R_{rep}$ is repetition rate of excitation process.
Therefore in addition to the availability of the necessary atomic states,
the optimal excitation method depends on the details of the experimental setup,
such as the time required to prepare the initial state as well as cavity QED parameters.

To estimate the time required to complete a single STIRAP photon generation,
we need to add the time necessary to prepare the initial state $|i\rangle$ as well as
the time for the adiabatic evolution and/or emission.
Adiabaticity requires that $\dot{\Omega}_P \ll g^2$,
and if $\Omega_P$ increases linearly over time $T$, $T \gg 1/g$.
In the strong coupling regime ($g \gg \kappa, \gamma$),
slow ramping of $\Omega_P$ guarantees that the quantum state of the system will be adiabatically transferred
from $|i,n=0\rangle$ to $|f,n=1\rangle$ without spontaneous emission where $n$ is the number of photons in the cavity mode.
The single photon then leaves the cavity with a rate of $\kappa$.
The Rabi rate $\Omega_P$ should remain at its maximum value until the single photon leaves the cavity.
Therefore with strong coupling, the total photon generation time includes 
$\mathcal{O}(1/\kappa)$ as well as time necessary for adiabatic process ($\gg 1/g$).
In the weak coupling regime, the cavity decay process will occur much faster than the adiabatic transfer,
and therefore the minimum time is mainly limited by $T \gg 1/g$.
Because Yb ion-cavity system lies in weak coupling regime,
STIRAP process will require much more time to extract a photon compared to fast excitation case.
In the rest of this paper, we will mainly consider fast excitation to compare the cavity result with high NA optics in the following section.

\section{Collection of photons with high numerical aperture optics}
\label{HighNASection}

\begin{figure}
	\begin{center}
		\includegraphics[width=8cm]{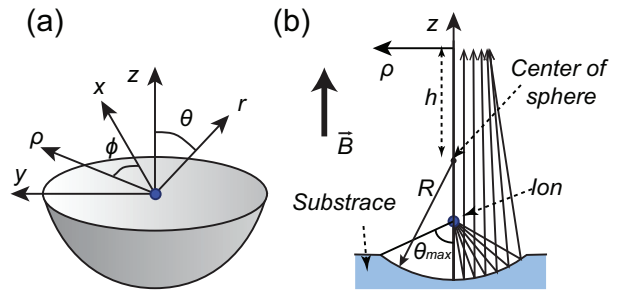} 
	\end{center}
	\caption{Definitions of coordinate systems and symbols with respect to spherical mirror.
			(a) Ion is located at the origin of coordinate system, and polar (cylindrical) coordinates are
			used to describe the optical field before (after) reflection. $\theta$ is the polar angle with respect to
			$\hat{z}$-axis, and $\phi$ is azimuthal angle from $\hat{x}$-axis.
			(b) $\theta_{max}$ is the maximum half opening angle corresponding to the mirror aperture.
			$h$ is a distance to the analysis plane from the center of sphere.
			Surface ion trap can be fabricated on the substrate area surronding mirror.
	}
	\label{Fig:CoordinateSystem}
\end{figure}
As an alternative method to collect light from a single trapped ion, 
we consider the use of high NA collection optics
whose geometric aberrations are well-corrected to achieve diffraction limited coupling performance into a single mode fiber.
Reflective optics is frequently used as an alternative to refractive optics to realize a large NA
\cite{StylusTrap,MEMSMirrorCharacterization,IonTrapWithSphericalMirror,SphericalMirrorWithCorrector,
ParabolicModeConverter,FiberCouplingByParabolicMirror,ParabolicMirrorGeometry}. 
A parabolic mirror can perfectly collimate light emitted from a point source, 
and has been studied extensively \cite{ParabolicModeConverter,FiberCouplingByParabolicMirror,ParabolicMirrorGeometry}.
In this section we will investigate the properties of light reflected by a spherical mirror
and numerically calculate the collection probability into a single mode fiber.
Although the analysis is presented for reflective optics,
the general conclusions apply to other approaches if similar NA and aberration control can be realized.

\subsection{Dipole emission and Gaussian mode}
\label{HighNAAnalysis}
Distinguishing light from $\pi$ and $\sigma$ transitions is essential for achieving high fidelities in remote entanglement protocols.
Therefore, the modes of photons from $\pi$ and $\sigma$ transitions have to be orthogonal after collection.
For large numerical aperture, this is only possible
if we choose $\vec{\it {B}}$ parallel to the optical axis $\hat{z}$.
The optical fields of photons emitted by the ion via the three possible transitions 
are given by \cite{FiberCouplingByParabolicMirror}
\begin{eqnarray}
\vec{E}_{\Delta l = 1,\Delta m=0} & = & \frac{ie^{ikr}}{r}\sqrt{\frac{3}{8\pi}}\sin\theta \hat{\theta},~~ \nonumber \\
\vec{E}_{\Delta l = 1,\Delta m=\pm1} & = & \frac{ie^{ikr}}{r}e^{\pm i\phi}\sqrt{\frac{3}{16\pi}}(\pm \cos\theta \hat{\theta}+i \hat{\phi}).
\label{Eq:OpticalFieldAtIon}
\end{eqnarray}
Once it is reflected by a curved surface,
spherical waves from an ion become paraxial waves, 
and the reflected optical field will be described in cylindrical coordinates shown in Figure \ref{Fig:CoordinateSystem} (a).
If the curved surface is parabolic and the trapped ion is located at the focus of paraboloid, 
all the reflected rays propagate parallel to the $\hat{z}$-axis and the traveling wavefronts are flat.

For a spherical mirror, most of the reflected rays propagate with radially varying phase delay 
and non-zero angle with respect to the $\hat{z}$-axis due to spherical aberration, even if the ion is trapped at the nominal focus 
(midpoint between the center of sphere and the vertex of the spherical surface).
In this case, rays cross each other and 
the optical field distribution in $\rho$-coordinate changes as it propagates along the $\hat{z}$-direction [Figure \ref{Fig:CoordinateSystem} (b)].
The probability of collecting a photon into a single mode fiber can be estimated 
by numerically calculating the mode matching factor [Eq. (\ref{Eq:ModeMatching})]
between the reflected optical field distribution ($ \vec{R}_{l,m}$) incident on the fiber 
and the Gaussian fiber mode ($ \vec{G}_{fiber}$) by setting $\vec{E} = \vec{R}_{l,m}$.

Luo {\it et al.}  \cite{FiberCouplingByParabolicMirror} analyzed the case of a parabolic mirror 
and calculated the overlap of the reflected mode with a Gaussian mode.
They showed that the collection probability of a photon emitted by the $\pi$-transition ($\Delta l=1,\Delta m=0$) 
vanishes due to the rotational cylindrical symmetry of polarization,
while the collection efficiency for $\sigma$-transition ($\Delta l=1,\Delta m=\pm 1$) approaches 50 \%
as the mirror aperture ($\rho_0$) becomes much larger than the focal length ($f$) of the parabolic mirror.
Provided the axis of single mode fiber is collinear with the quantization axis,
a single mode fiber can be used as transition-selective filter which maps photons generated by $\sigma^+$ and $\sigma^-$-transitions
into two orthogonal circular polarizations ($(\hat{x}\mp i\hat{y})/\sqrt{2}$ ) respectively, 
and rejects photons generated by the $\pi$ transition.
For the frequency qubit shown in Fig. \ref{Fig:EntanglementScheme} (d), it would be interesting to collect photons from $\pi$-transitions
and reject  $\sigma$ transition photons \cite{QuantumTeleportationExperiment}.
We can implement this orthogonal filter by adding an additional $\phi$-dependent phase ($e^{i\phi}$) to the optical field
after it is reflected by the mirror [Eq. (33) and (35) of Ref. \cite{FiberCouplingByParabolicMirror}].
A phase that depends on the azimuthal angle can be added
using computer generated holograms \cite{LGmodeByHolograms1,LGmodeByHolograms2},
diffractive optics \cite{LGmodeByDiffraction1,LGmodeByDiffraction2}, 
phase plates \cite{LGmodeByPhasePlate1,LGmodeByPhasePlate2},
or spatial light modulators \cite{LGmodeBySpatialLightModulator1,LGmodeBySpatialLightModulator2}.
Therefore we assume that whenever we intend to collect light from $\pi$-transitions, 
an additional phase factor $e^{i\phi}$ is added to the (near-) collimated field.

\begin{figure}
	\begin{center}
		\includegraphics[width=8cm]{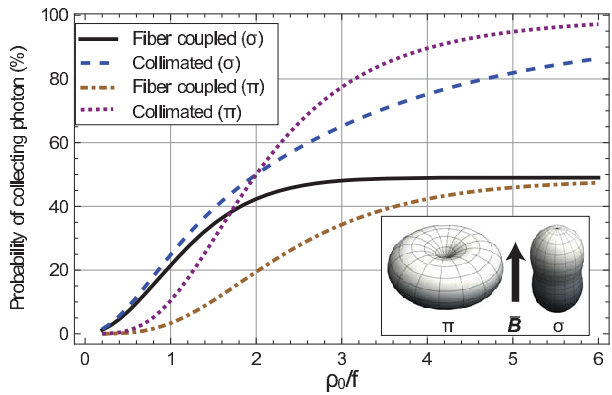} 
	\end{center}
	\caption{Maximum collection probability of photons emitted by a $\sigma$-transition ($\Delta l=1,\Delta m=\pm 1$) and a $\pi$-transition ($\Delta l=1,\Delta m=0$).
		For the $\pi$-transition calculation, we assume that phase $e^{i\phi}$ is added to the collimated field.
		We also plotted the amount of light collimated by parabolic mirror before being collected by single mode fiber.
		$\rho_0$ is the radius of mirror aperture and $f$ is the focal length of the parabolic mirror.
		Inset: Spherical probability distribution of photons emitted by different transitions 
		with respect to the quantization axis.
		Distance from the origin is proportional to the probability of photons propagating in the corresponding angle.
	}
	\label{Fig:ParabolaProbability}
\end{figure}

The collection probabilities of light from a single ion onto a large area detector (referred to as ``collimated case'')
and into a single mode fiber (``fiber coupled case'') are shown in Figure \ref{Fig:ParabolaProbability}
as a function of the mirror aperture radius ($\rho_0$) 
for both $\sigma$- and $\pi$-transitions.
When $\rho_0$ is smaller than $f$, 
the fiber coupling efficiency of $\sigma$-transition photons is much larger than that of $\pi$-transition photons
due to the dipole emission pattern [inset in Figure \ref{Fig:ParabolaProbability}].
When $\rho_0$ is much larger than $f$, 
spatial overlap with fiber mode becomes less sensitive to the details of the emission pattern
but the spatially varying polarization distribution limits the overall coupling efficiencies of both transitions near 50 \%.

\begin{figure}
	\begin{center}
		\includegraphics[width=8cm]{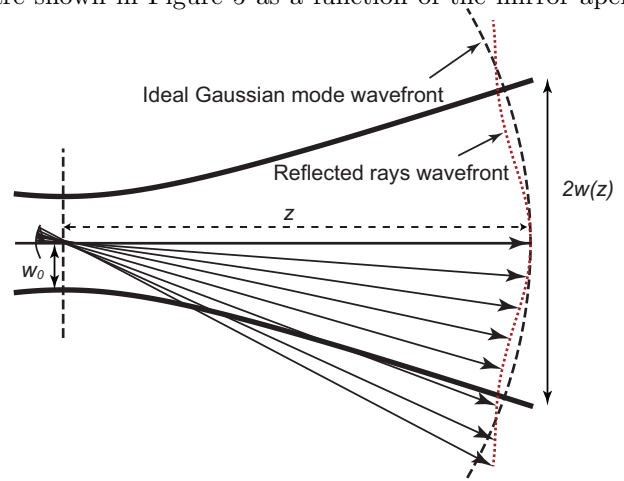} 
	\end{center}
	\caption{Matching collimated ray distribution to Gaussian beam with finite wavefront radius of curvature.
			For clarity, figures are not drawn to scale.}
	\label{Fig:GaussianProfile}
\end{figure}

The integration of an ion trap with large aperture parabolic mirrors has been considered \cite{StylusTrap},
In this work, we will concentrate our analysis on the collection probability 
one can achieve with a spherical mirror,
which can be integrated with a surface trap, as shown schematically in Figure \ref{Fig:CoordinateSystem} (b) \cite{MEMSMirrorCharacterization}.
As spherical mirrors feature substantial aberrations at large NA, this analysis also provides strategies for effectively managing aberrations present
in most practical collection optics.
When the ion is trapped at the nominal focus of the spherical mirror, 
the rays collimated by a spherical mirror are not perfectly parallel to the optical axis 
but converge initially and eventually diverge after crossing the optical axis (Figure \ref{Fig:GaussianProfile}).
The resulting far-field wavefronts are close to spherical with residual optical path difference (OPD)
as shown in Fig. \ref{Fig:GaussianProfile}.
To estimate the probability of coupling this mode to an optical fiber,
we calculate the mode overlap to a Gaussian mode with spherical wavefronts:
\begin{widetext}
\begin{equation}
\vec{G}(\rho,\phi,z)=\frac{1}{w}\sqrt{\frac{2}{\pi}} \exp\left(i\psi\right) \exp\left(-\frac{\rho^2}{w^2(z)}\right) \exp\left(\frac{ik\rho^2}{2 R(z)} \right) (\alpha\hat{x}+\beta\hat{y}),
\end{equation}
\end{widetext}
where $R(z)$ is the radius of curvarture (RoC) of the wavefront, $w(z)$ is the beam radius at the measurement plane, 
and $\psi = \arctan(-z/2z_R)$ is Gouy phase.
The residual OPD between the wavefront of collimated light field 
and the spherical wavefronts of the Gaussian mode 
limits the maximum coupling efficiency one can achieve with this method. The coupling efficiency can be recovered to values similar to
the ideal parabolic case if the residual OPD is reduced to below the diffraction limit.

\subsection{Design examples}
\label{HighNAExample}
\begin{figure}
	\begin{center}
		\includegraphics[width=8cm]{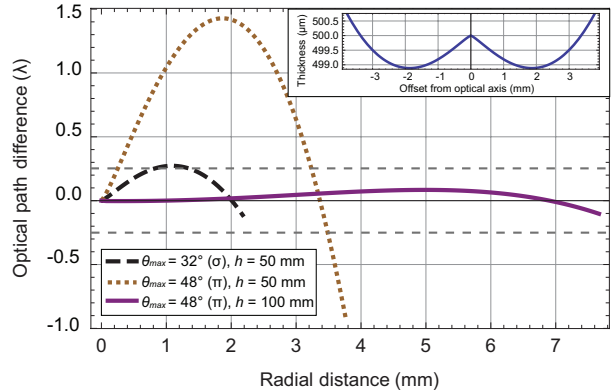} 
	\end{center}
	\caption{Example of residual OPD expressed in units of wavelength ($\lambda=369.5$~nm).
	The RoC of matching wavefront is chosen to maximize the mode overlap rather than minimizing the maximum OPD.
	According to Rayleigh wavefront criteria \cite{RayleighWavefrontCriterion}, maximum OPD should be less than $\lambda/4$ to avoid the effect of non-zero OPD
	and this range is marked by two dashed lines.
	The dark dashed line and dotted line shows the residual OPD (measured at a location $h$ = 50~mm)
	for the rays reflected by a spherical mirror ($R$ = 160~$\mu$m)
	with $\theta_{max}$ = 32$^\circ$ and $\theta_{max}$ = 48$^\circ$, respectively.
	In the inset, thickness variation of phase plate is plotted, designed to compensate OPD shown by the dotted line. 
	500$\mu m$-thick fused silica substrate is assumed in this design.
	With this phase correcting plate placed at $h$ = 50~mm, residual OPD at $h$ = 50~mm is dramatically suppressed.
	Solid line is OPD of phase corrected wavefront measured at $h$ = 100~mm,
	which shows that OPD barely changes for additional propagation once OPD is compensated.
	}
	\label{Fig:OPDPlot}
\end{figure}

The residual OPD (expressed in units of wavelength $\lambda$ = 369.5~nm)
of the rays reflecting off a micromirror with RoC $R$ = 160~$\mu$m,
measured at a location $h$ = 50~mm 
compared to a best-matching Gaussian mode is plotted in Figure \ref{Fig:OPDPlot}.
The ion is assumed to be located at the focal point of the spherical mirror.
The two horizontal gray dashed lines show the limits of Rayleigh criterion representing the diffraction limit (OPD $\le \pm \lambda/4$),
below which residual OPD does not lead to reduced coupling efficiency \cite{RayleighWavefrontCriterion}.
The black dashed line shows the residual OPD for a spherical micromirror with opening angle ($\theta_{max}$) of 32$^\circ$,
where 10.5 \% of $\sigma$-transition light falls within the aperture and is collimated by the mirror.
In this case, the OPD remains within the Rayleigh criterion, and 
a numerical calculation estimates that 6.2 \% of photons from $\sigma$-transitions can be matched to a single mode fiber.
In contrast, the light field generated by the $\pi$-transition has little overlap with a Gaussian mode for $\theta_{max}=32^\circ$
due to the emission pattern.

When $\theta_{max}$ is extended to $48^\circ$ to increase the amount of collimated light from $\pi$-transition to 7.3 \%,
the maximum OPD becomes larger than $\lambda$ due to spherical aberrations (dotted line in Figure \ref{Fig:OPDPlot}) and the maximum coupling efficiency is still limited to 0.4 \%.
The residual OPD must be compensated in order to recover the fiber coupling efficiency.
If the distribution of the OPD is known, one can utilize a phase plate \cite{LGmodeByPhasePlate1,LGmodeByPhasePlate2}
or a spatial light modulator \cite{LGmodeBySpatialLightModulator1,LGmodeBySpatialLightModulator2}
to introduce compensating OPD to cancel out the residual OPD.
Inset of Figure \ref{Fig:OPDPlot} shows an example of a phase plate 
which can be fabricated on a 500~$\mu$m-thick fused silica substrate to compensate the residual OPD for the $\pi$-transition case.
Using this phase corrector plate, the coupling efficiency into the fiber mode can be enhanced to 3.0 \%.
We used $h$ = 50~mm as a typical location to place phase correcting optics, but
the general conclusion is independent of where the mode matching is calculated.
We also explored optimization of ion height along the optical axis, and concluded that no substantial improvement can be achieved.

\begin{figure}
	\begin{center}
		\includegraphics[width=8cm]{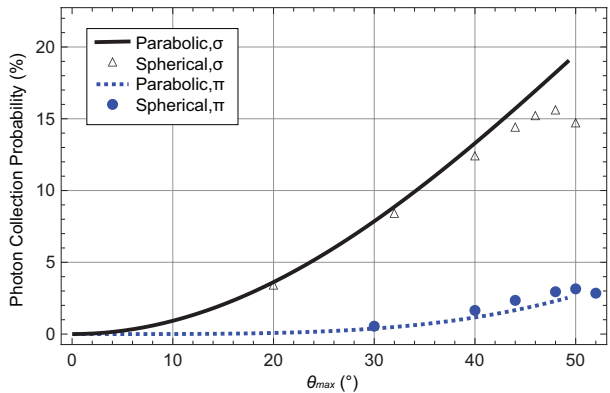} 
	\end{center}
	\caption{Fiber coupling efficiency of a photon emitted by $\sigma$- and $\pi$-transition using a spherical mirror with a phase corrector plate when $R=160\mu m$.
	We also plotted fiber coupling efficiency by parabolic mirror for comparison.
	This plot shows that most of aberration of spherical mirror can be corrected by phase plate when $\theta_{max}< 50^\circ$.}
	\label{Fig:TotalProbability}
\end{figure}

Figure \ref{Fig:TotalProbability} shows the probability of coupling emitted photons into a single mode fiber
using a spherical mirror and an optimized phase corrector plate for each value of $\theta_{max}$.
With optimum phase correction, the collection probability as a function of the opening angle $\theta_{max}$ of a spherical mirror 
grows similar to that of a parabolic mirror
until $\theta_{max}$ reaches $\sim 50^\circ$ where matching of slightly diverging rays to 
a Gaussian mode starts to fail.
Beyond this limit, we need to correct the angle of collimated rays in addition to the phase
which is beyond the scope of this paper.
Likewise, $\pi$-transition coupling efficiency could be improved to over 3.0 \% from 0.4 \%,
approaching the parabolic mirror limit if all aberrations are properly compensated.

Use of a microfabricated mirror integrated under a surface trap has been proposed
for efficient collection of emitted photons for qubit state detection \cite{MEMSMirrorCharacterization}.
In that approach, however, the aberrations of the microfabricated mirror cannot be controlled well enough to produce a flat wavefront.
One important advantage of using a microfabricated mirror compared to a conventional macroscopic mirror
\cite{SphericalMirrorWithCorrector} is that the geometric aberrations scale
as the dimensions of the optics are reduced \cite{MultiscaleOptics}.
The residual OPD is reduced while the wavelength ($\lambda$) remains the same eventually achieving diffraction limit.
Table \ref{Tab:collectionEfficiency} summarizes photon coupling efficiency
using a single spherical mirror with the same large NA ($\theta_{max}$ = 48$^\circ$) without any additional aberration correction.
Last column for RoC = 16~$\mu$m shows that there will be a significant improvement 
if the dimensions of the mirror and the position of the ion are shrunk by a factor of 10.
For such micro optics, the detailed shape of the reflector has little impact on the collection efficiency.
On the other hand, if the dimensions of mirror are increased by a factor of 10, 
the collection probability will vanish unless there is proper phase compensating optics (first column).
Therefore the microfabricated mirror with RoC $\ll$ 100~$\mu$m can increase the fiber collection efficiency drastically.
However due to technical limitations of the ion height, it is impractical to use mirrors with RoC much smaller than $\sim$ 100~$\mu$m at present day.

\begin{table}
\begin{tabular}{c|c|c|c|c|c|c} \hline
Mirror RoC ($\mu$m)&\multicolumn{2}{c|}{1600} & \multicolumn{2}{c|}{160} & \multicolumn{2}{c}{16}\\ \hline
Transition & $\sigma$ & $\pi$ & $\sigma$ & $\pi$ & $\sigma$ & $\pi$ \\ \hline
Collimated light by mirror (\%) & 21.2 & 7.3 & 21.2 & 7.3 & 21.2 & 7.3 \\
Fiber coupling probability (\%) & $<$0.1 & $<$0.1 & 6.2 & 0.4 & 15.8 & 2.8 \\ \hline
\end{tabular}
\caption{Estimation of fiber coupling efficiency of $\sigma$ and $\pi$ transition light by ideal spherical mirrors of different dimensions.
	$\theta_{max}$ is set to 48$^\circ$.
	In our simulation, we added an ideal hyperbolic lens in the beam path, which can transform the matched Gaussian mode into UV fiber mode ($w_0 \sim$ 1.54~$\mu$m)
	and performed full ray tracing from the ion position to the fiber plane to evaluate the mode overlap integral.}

\label{Tab:collectionEfficiency}
\end{table}

\subsection{Alternative approaches}
\label{alternatives}
The analysis in this section shows that efficient coupling of light into a single mode fiber can be achieved
provided one can correct the aberration of a large NA collection optics to conform to the wavefront of a Gaussian beam, to below the diffraction limit.
Practical means to realize such high NA optics with well-corrected aberration include
(1) design of multi-element refractive lenses,
(2) fabrication of paraboloidal mirrors (using e.g. diamond turning technology),
(3) spherical mirror with corrector plates to compensate for the residual aberration,
and (4) Fresnel reflectors or diffractive optical components.

Fresnel lenses have been investigated as a way to collect light with large NA without spherical aberration \cite{FresnelLensCalculation,FresnelLensImaging}
because it adds the necessary phase shift to a continuous wavefront.
Using the similar design principle,
we can consider integrating Fresnel reflector with an ion trap
in place of the micro mirror shown in Figure \ref{Fig:CoordinateSystem} (b).
One critical consideration for Fresnel diffractive optical elements
is the diffraction efficiency that depends on the grating periods \cite{FresnelLensCalculation}.
The Fresnel reflector deserves more careful analysis for use in
aberratioin-free fiber coupling applications.

\section{Application to remote entanglement generation}
\label{Discussion}
Based on our estimation of photon coupling efficiency into a single mode fiber,
we can calculate the expected entangled ion pair generation rate considering practical experimental constraints.
Table \ref{Tab:GenerationRate} compares the expected entanglement generation rate
of various approaches with existing experimental results.
Matsukevich {\it et al.} \cite{RemoteEntanglementPolarization} generated entangled ion pairs
by using polarization qubit [Figure \ref{Fig:EntanglementScheme} (b)].
Because state preparation step requires only optical pumping into $^2S_{1/2}|0,0\rangle$
without any qubit rotation \cite{YbSystem}, the same procedure can be repeated at 520 kHz repetition rate,
and the entangled pair generation rate is analyzed in the first column of Table \ref{Tab:GenerationRate}.
When the remote entanglement is established by coincident detection of a photon pair,
the ionic qubit is encoded in the Zeeman substates $^2S_{1/2}|1,\pm1\rangle$
which are susceptible to magnetic field fluctuation and therefore 
these two states should be transferred to atomic clock states $^2S_{1/2}|\pm1,0\rangle$ by microwave transitions or Raman transitions.
The photonic polarization qubit is also susceptible to decoherence from birefringence in the vacuum window, downstream optics and single mode fibers.
Therefore this method requires careful control of potential decoherence mechanisms to achieve high fidelity.
Olmschenk {\it et al.} \cite{QuantumTeleportationExperiment} 
used the first-order magnetic field-insensitive hyperfine clock state qubit 
accompanied by $\pi$ transition as shown in Figure \ref{Fig:EntanglementScheme} (d),
and collected emitted photons from a direction orthogonal to the quantization axis
to maximize collection of photons generated by $\pi$ transition [inset of Figure \ref{Fig:ParabolaProbability}].
To generate ion-photon entangled state, the ion should be prepared in a superposition state before excitation,
which requires rotation of clock states using microwave transition and limits the overall repetition rate to 75 kHz.
The analysis of this experiment is listed in the second column of Table \ref{Tab:GenerationRate}.
The limitation of this collection geometry is the increased overlap between different polarizations from $\sigma$ and $\pi$ transitions,
which reduces fidelity as NA is increased \cite{QuantumTeleportationExperiment}.


\begin{table*}
\begin{tabular}{c|c|c|c|c|c|c} \hline
Numerical aperture &\multicolumn{2}{c|}{0.23} & \multicolumn{3}{c|}{0.6} & cavity\\ \hline
Optic axis vs. B-field & $\parallel$ & $\perp$ & $\parallel$ & $\parallel$ & $\parallel$ & $\parallel$ \\
Type of atomic transition & $\sigma$ & $\pi$ & $\sigma$ & $\pi$ & $\sigma$ & $\sigma$ \\
Photonic qubit & Pol. & Freq. & Pol. & Freq. & Freq. & Pol. \\ \hline
Detector efficiency & 0.2 & 0.2 & 0.3 & 0.3 & 0.3 & 0.3 \\ 
Decay via right transition & 2/3 & 1/3 & 2/3 & 1/3 & 2/3 & n.a. \\ 
Collected fraction & 0.0198 & 0.0198 & 0.136 & 0.028 & 0.136 & 0.337 \\ 
Mode overlap & 0.82 & 0.82 &0.85 & 0.32 & 0.85 & 0.95 \\
Misc. efficiency & 0.26 & 0.25 & 0.63 & 0.62 & 0.63 & 0.63 \\ \hline
Single Photon collection efficiency (\%) & 0.0554 & 0.0269& 1.46 & 0.0547& 1.46 & 6.04 \\ \hline
Bell state identification & 0.25 & 0.25 & 0.5 & 0.5 & 0.5 & 0.5\\ \hline
Coincidence efficiency & $7.7\times10^{-8}$ & $1.8\times 10^{-8}$ & $1.1\times10^{-4}$ & $1.5\times10^{-7}$ & $1.1\times10^{-4}$ & $1.8\times10^{-3}$ \\ \hline
Repetition rate (kHz) & 520 & 75 & 500 & 75 & 75 & 500 \\ \hline
Entanglement rate (Hz) & 0.04 & 0.0014 & 53 & 0.011 & 8 & 913 \\ \hline
\end{tabular}
\caption{Comparison of entanglement generation rate based on different setup. 
Theoretically maximum two types of Bell states encoded in a photon pair can be
unambiguously identified from the others, and we include this identification efficiency on the 5th row.
When atom decays to the ground state, some of the decay channels are undesired
and we included the fraction of desired decay channel in 7th row.
Collected fraction corresponds to the fraction of collimated light 
when the photons are collected with reflective or refractive optics
and $r_t \cdot P_{cavity}$ for the case of the cavity system.
Experimental condition of each column is described in the main text.}
\label{Tab:GenerationRate}
\end{table*}

From the third  to the fifth column in Table \ref{Tab:GenerationRate} shows the expected entangled ion pair generation rates
when the numerical aperture is increased to 0.6 using either reflective optics or refractive optics
discussed in Section \ref{HighNASection}.
We considered several practical improvements that can be made over the existing experiments.
Although previous experiments \cite{RemoteEntanglement,RemoteEntanglementPolarization,QuantumTeleportationExperiment}
detected only the singlet state out of four Bell states, 
one of triplet states can also be distinguished from the other two.
Therefore we assume that probability of Bell state measurement is 1/2 rather than 1/4.
In previous experiments, various losses in photon collection efficiency
caused by non-ideal optical components further reduced the overall efficiency by a factor of $\sim$0.25.
We estimate that we can use better optical components to increase this efficiency to more than 60 \% based on realistic analysis.
Third column analyzes experimental setup similar to the first column \cite{RemoteEntanglementPolarization}
when the numerical aperture is increased to 0.6.
This setup can achieve the highest expected generation rate among the three cases,
but requires decoherence control similar to the first column setup.
Forth column lists values we can achieve using similar setup to the second column \cite{QuantumTeleportationExperiment},
but we assume that optic axis of collection optics is parallel to the quantization axis 
compared to orthogonal setup used in the second column.
As was discussed in section \ref{HighNASection}, if a phase plate is used in the parallel configuration,
only the $\pi$ transition can be coupled into a single mode fiber and the fidelity will remain high even with very large numerical aperture.
However, NA of 0.6 is not large enough to show significant improvement 
due to non-ideal emission pattern of $\pi$-transition in this collection direction.
Fifth column estimates the generation rate which can be achieved 
with a scheme in Figure \ref{Fig:EntanglementScheme} (c) \cite{RemoteEntanglement}.
This approach has advantages of high fidelity due to frequency qubit
and $\sigma$ transition which allows high collection efficiency.
However, to achieve near unity excitation probability, the initial state should be prepared to
either a superposition of the clock states or one of $^2S_{1/2}|1,\pm1\rangle$, which 
limits the repetition rate around 75 kHz.

The last column shows the entanglement generation rate which can be achieved by an optical cavity 
discussed in Section \ref{CavitySection}.
In schemes using optical cavities, the FSR of cavity should be matched to 
either 12.6 GHz or 14.7 GHz to collect photons encoding frequency qubits.
These constraints lead to minimum cavity length of $\sim$10~mm,
precision control of RoC of the spherical mirror down to $\mu$m level,
and an upper limit on the cavity finesse of $\sim$ 700 to cover 19.6 MHz linewidth of the $^2P_{1/2}$ state.
Therefore frequency qubit is not compatible with optical cavity schemes.
However, we can achieve highest entanglement generation rate
compared to other approaches using polarization qubit or time-bin qubit.
In this case, the remaining challenge will be the minimization of decoherence 
similar to all the other schemes involving the polarization qubit.

Based on the above analysis, collection of photons generated by $\sigma$-transition
along quantization axis is the optimal choice for both the large NA system and the cavity system.
Cavity system prefers polarization qubit as the photonic qubit of choice,
but it can achieve the highest entanglement generation rate.
For large NA system, frequency qubit has lower generation rate than polarization qubit
mainly due to slow initial state preparation.
In principle, this limitation can be overcome with fast microwave transition or Raman transition.
A custom lens system designed for optimal chamber configuration can achieve a NA of more than 0.6
with diffraction-limited performance.
Even though the maximum collection probability with free space optics is smaller,
free space optics has the advantage that it can be realized with off-the-shelf technology today. Therefore,
$\sigma$-transition-based frequency qubit system with large numerical aperture lens
can be a good compromise between high fidelity and high throughput.

\section{Conclusions}
\label{Conclusions}
In this paper, we discussed various ways to enhance collection of photons emitted by a single atomic ion into a single-mode optical fiber.
Based on this analysis, we conclude that the generation rate of entangled ion pairs through entanglement swapping of photons can be dramatically
enhanced beyond the current values using either a high NA collection optics or a carefully designed optical cavity. Such enhancement is expected
to enable the realization of scalable quantum multi-processor and quantum repeaters based on nodes of trapped ions connected via photonic network.

This research was funded by Office of the Director of National Intelligence  and Intelligence Advanced Research Projects Activity through Army Research Office.

\appendix
\section{Cavity with two concave mirrors}
\label{Appendix:A}
In this Appendix, we estimate the maximum collection efficiency 
one can achieve with a cavity made of two curved mirrors with an ion trapped in the center.
Compared to the fiber cavity, ion can be trapped at the beam waist of the cavity mode,
and we can estimate the lower bound and the upper bound of $P_{cavity}$.
In the near-concentric limit, we estimate the optimal $w_0$ to be 
$0 < w_0 < \sqrt{\lambda/\pi} \sqrt[4]{\text{RoC}\cdot \lambda/2 }$,
which corresponds to $1 > 1/(1+(w_0/r_c)^2) > 0.87$.
Combined with $1/(1+\text{2RoC}/l_c) \approx 0.53$ which is possible with symmetric geometry,
we can estimate that $0.53 > P_{cavity} > 0.46$ which is comparable to $P_{cavity} \approx$ 0.41 of the fiber cavity.
Mode matching has to be performed to couple these photons into a single mode fiber, and it is not easy to integrate this geometry with surface traps.

\bibliography{collection}

\end{document}